\documentclass[useAMS,usenatbib]{mnras}
\usepackage{graphicx} 
\usepackage{subcaption}
\usepackage{amsmath} 
\usepackage{xcolor}
\usepackage{float}
\usepackage{caption}
\usepackage{placeins}
\usepackage{lipsum}
\usepackage{multicol}
\usepackage{soul}
\usepackage{amssymb}
\usepackage{ragged2e}
\usepackage{threeparttable}

\def\refjnl#1{{\rm#1}}
\def\aj{\refjnl{AJ}}                   
\def\araa{\refjnl{ARA\&A}}             
\def\apj{\refjnl{ApJ}}                 
\def\apjl{\refjnl{ApJ}}                
\def\apjs{\refjnl{ApJS}}               
\def\ao{\refjnl{Appl.~Opt.}}           
\def\aap{\refjnl{A\&A}}                
\def\baas{\refjnl{BAAS}}               
\def\jcap{\refjnl{J. Cosmology Astropart. Phys.}} 
\def\mnras{\refjnl{MNRAS}}             
\def\pasa{\refjnl{PASA}}               
\def\pasj{\refjnl{PASJ}}               
\def\nat{\refjnl{Nature}}              
\def\physrep{\refjnl{Phys.~Rep.}}   

\def\be{\begin{equation}} 
\def\ee{\end{equation}}

\def\gsim{\lower.5ex\hbox{\gtsima}} 
\def\lsim{\lower.5ex\hbox{\ltsima}} \def\gtsima{$\; \buildrel > \over 
\sim \;$} \def\ltsima{$\; \buildrel < \over \sim \;$} \def\prosima{$\; 
\buildrel \propto \over \sim \;$} \def\gsim{\lower.5ex\hbox{\gtsima}} 
\def\lsim{\lower.5ex\hbox{\ltsima}} 
\def\simgt{\lower.5ex\hbox{\gtsima}} 
\def\simlt{\lower.5ex\hbox{\ltsima}} 
\def\simpr{\lower.5ex\hbox{\prosima}} \def\la{\lsim} \def\ga{\gsim} 
  
 \def\gtsima{$\; \buildrel > \over \sim \;$} 
\def\ltsima{$\; \buildrel < \over \sim \;$} 
\def\gsim{\lower.5ex\hbox{\gtsima}} 
\def\lsim{\lower.5ex\hbox{\ltsima}} 
\def\simgt{\lower.5ex\hbox{\gtsima}} 
\def\simlt{\lower.5ex\hbox{\ltsima}} 
\def\simpr{\lower.5ex\hbox{\prosima}} 
\def\la{\lsim} 
\def\ga{\gsim}

\def\E3{{\cal E}_{\rm g}^{III}}

\def\mbh{\rm M_{\rm BH}}
\def\lbh{\rm L_{\rm BH}}
\def\lUV{\rm L_{\rm UV}}

\def\Msun{\, \rm M_\odot}

\def\zsun{\rm Z_\odot}
\def\M*{M_*}
\def\Z*{Z_*}
\def\L*{L_*}
\def\muv{\rm M_{UV}}

\newcommand{\fedd}{\,{f_{\rm Edd}}}

\title[Interpreting the $z \sim 13$ Drop-out Sources]{Are the Newly-Discovered $z \sim 13$ Drop-out Sources Starburst Galaxies or Quasars?}

\author[F. Pacucci et al.]{
Fabio Pacucci$^{1,2}$\thanks{fabio.pacucci@cfa.harvard.edu},
Pratika Dayal$^{3}$\thanks{p.dayal@rug.nl},
Yuichi Harikane$^{4,5}$,
Akio K. Inoue$^{6,7}$ \&
Abraham Loeb$^{1,2}$
\\
$^{1}$Center for Astrophysics $\vert$ Harvard \& Smithsonian, Cambridge, MA 02138, USA\\
$^{2}$Black Hole Initiative, Harvard University, Cambridge, MA 02138, USA\\
$^{3}$Kapteyn Astronomical Institute, University of Groningen, P.O. Box 800, 9700 AV Groningen, The Netherlands\\
$^{4}$Institute for Cosmic Ray Research, The University of Tokyo, 5-1-5 Kashiwanoha, Kashiwa, Chiba 277-8582, Japan\\
$^{5}$Department of Physics and Astronomy, University College London, Gower Street, London WC1E 6BT, UK\\
$^{6}$Department of Physics, School of Advanced Science and Engineering, Faculty of Science and Engineering, \\ Waseda University, 3-4-1 Okubo, Shinjuku, Tokyo 169-8555, Japan\\
$^{7}$Waseda Research Institute for Science and Engineering, Faculty of Science and Engineering, \\ \:Waseda University, 3-4-1 Okubo, Shinjuku, Tokyo 169-8555, Japan
}

\date{\today}

\pubyear{2021}

\begin{document}
\label{firstpage}
\pagerange{\pageref{firstpage}--\pageref{lastpage}}
\maketitle

\begin{abstract}
The detection of two $z\sim 13$ galaxy candidates has opened a new window on galaxy formation at an era only $330$ Myr after the Big Bang. Here, we investigate the physical nature of these sources: are we witnessing star forming galaxies or quasars at such early epochs? If powered by star formation, the observed ultraviolet (UV) luminosities and number densities can be jointly explained if: (i) these galaxies are extreme star-formers with star formation rates $5-24\times$ higher than those expected from extrapolations of average lower-redshift relations; (ii) the star formation efficiency increases with halo mass and is countered by increasing dust attenuation from $z \sim 10-5$; (iii) they form stars with an extremely top-heavy initial mass function. The quasar hypothesis is also plausible, with the UV luminosity produced by black holes of $\sim 10^8 \Msun$ accreting at or slightly above the Eddington rate ($f_{\rm Edd}\sim 1.0$). This black hole mass at $z\sim13$ would require very challenging, but not implausible, growth parameters. If spectroscopically confirmed, these two sources will represent a remarkable laboratory to study the Universe at previously inaccessible redshifts.
\end{abstract}



\section{Introduction}
\label{sec:introduction}
The Universe began in the Big Bang with very simple initial conditions, which can be neatly summarised by the standard cosmological parameters (see, e.g., \citealt{WMAP_2013, Planck_2018}). Once the physics of baryons started to have a more profound influence on the cosmic evolution, the Universe became increasingly complex with the start of galaxy formation and reionization \citep[see e.g.][]{BL01, dayal2018}. In order to bridge the gap between the high-$z$ and the complex local Universe, the detection of ever-higher redshift sources is essential to understand when and how the first stars and black holes formed (see, e.g., \citealt{Woods_2019}). 

Currently, the highest-redshift source ever detected with a spectroscopic confirmation is a galaxy at $z=10.957$ \citep{jiang2021} with a stellar mass of $\rm M_{\star} \approx 10^9 \Msun$ \citep{Oesch_2016}, while the highest-redshift quasar is a galaxy hosting a black hole of mass $\mbh \approx 1.6\times 10^9 \Msun$ at $z = 7.642$ \citep{Wang_2021}. 
Current predictions suggest that the highest-redshift quasar, defined as a black hole of mass $> 10^9\Msun$, could be detected by upcoming surveys in the redshift range $z=9-12$ \citep{Euclid_2019, Fan_2019BAAS}. These future detections, with facilities including Euclid and the Rubin Observatory, will have profound repercussions on our ability to constrain the growth parameters of early populations of black holes \citep{Pacucci_2020, Pacucci_2022}.

Recently, \cite{harikane2022} presented the detection of two H-band drop-out Lyman-Break Galaxy (LBG) candidates at $z\sim 13$ (HD1 and HD2). Although also explicable at $z \sim 4$, the $z\sim 13$ solutions to photometric fitting offer significantly lower $\chi^2$ statistics. While not confirmed by spectroscopy, one source is backed by a $4\sigma$ signal of the [OIII]$88\mu m$ line at $z = 13.27$ with the Atacama Large Millimeter Array (ALMA), further supporting the $z \sim 13$ estimate. Although a confirmation of the redshift of these two sources will require spectroscopic observations, this claim paves the way to performing a deeper search into existing data to look for very high-$z$ sources that could have been potentially missed thus far. \cite{harikane2022} also propose surveys to look for more of such sources with forthcoming facilities including the James Webb Space Telescope (JWST) and the Nancy Grace Roman Space Telescope (NGRST). 

These two sources appear extremely ultraviolet (UV) bright (with $\muv \sim -23.5$), and were tentatively explained in \cite{harikane2022} as starburst galaxies, AGN or sources with a top-heavy IMF. In this Letter, we further investigate the physical viability of the starburst galaxy hypothesis, while also proposing a quasar hypothesis, i.e., these two UV-bright galaxies might be accreting super-massive black holes at $z\sim 13$. Here, we do not consider the possibility that the sources are lensed because they are spatially isolated from other nearby sources.
We use the \cite{Planck_2018} cosmology as a reference.

\section{The Star-forming galaxy hypothesis}
\label{sec:starburst}

\begin{figure}
\center{\includegraphics[scale=0.8]{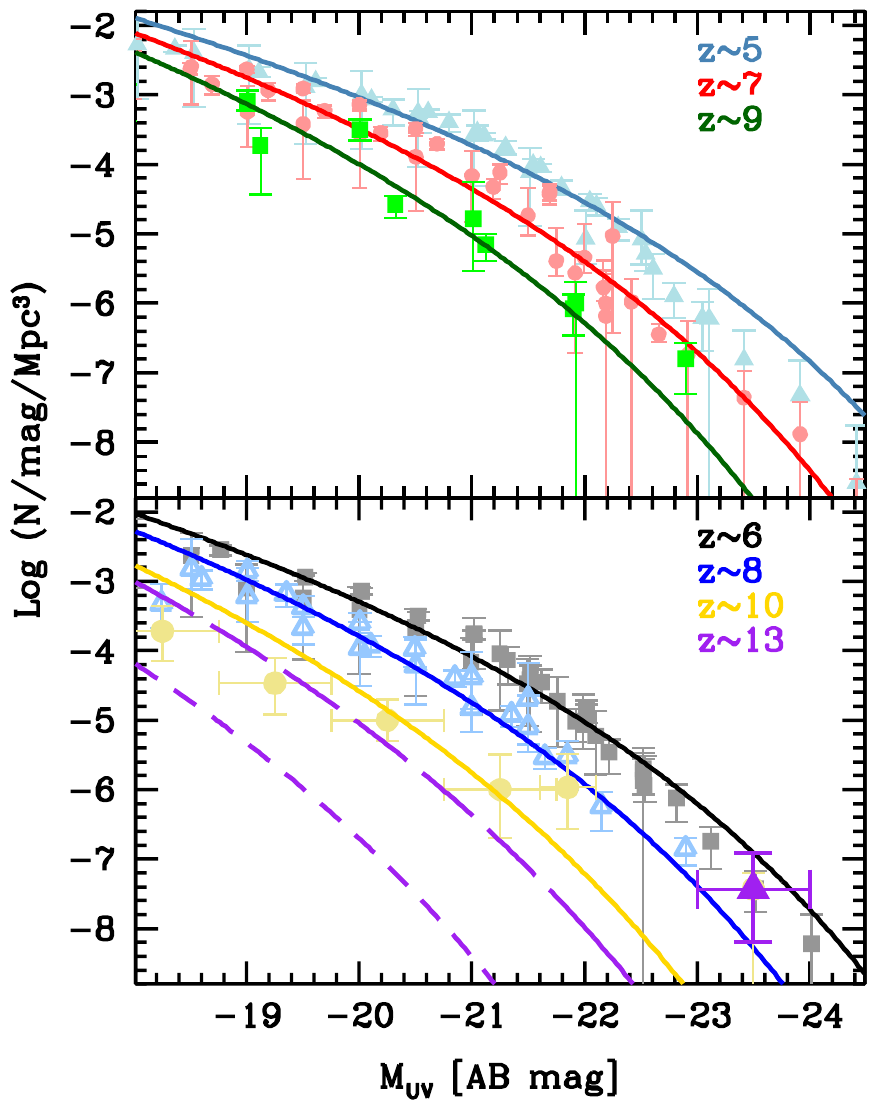}}
    \caption{The evolving UV LF at $z \sim 5-7$ ({\it upper panel}) and $z \sim 8-13$ ({\it lower panel}). Points show observational data: $z \sim 5$ \citep[filled cyan triangles;][]{finkelstein2015, bouwens2021, Harikane_2021}, $z \sim 6, 7$ \citep[empty gray squares and empty red circles, respectively;][]{bowler2015,finkelstein2015, bouwens2021, Harikane_2021}, $z \sim 8$ \citep[empty blue triangles;][]{finkelstein2015, ishigaki2018, bowler2020, bouwens2021}, $z \sim 9$ \citep[filled green squares;][]{ishigaki2018, bowler2020, bouwens2021}, $z \sim 10$ \citep[filled yellow circles;][]{oesch2018, bowler2020, Harikane_2021} and $z \sim 13$ \citep[filled purple triangle;][]{harikane2022}. In both panels, the solid lines show theoretical UV LFs obtained by tuning the value of the star formation efficiency to the data at each redshift (see Table. \ref{tab:sfe}). The short- and long-dashed lines show the UV LF at $z \sim 13$ assuming that the star formation efficiency saturates to a value of $\sim 0.5\%$ at $z \gtrsim 9$ (case 1) and using the redshift evolution of the star formation efficiency inferred from $z \sim 5-9$ data (Eqn. \ref{sfe_fnz}; case 2), respectively. }
    \label{fig:uvlf}
\end{figure}

In this section, our key aim is to check whether the two $z\sim 13$ drop-out sources detected by \cite{harikane2022} are compatible with the extrapolation of lower-$z$ ($z \sim 5-10$) star-forming galaxy data, i.e. if they could be explained as star-forming galaxies. We start from the Sheth-Tormen halo mass function \citep[HMF;][]{sheth-tormen2002} and assume each halo to contain a gas mass that is linked to the halo mass ($\mathrm{M_h}$) through the cosmological baryon-to-dark matter ratio. Assuming bursty star formation, the UV luminosity ($\lUV$) of a halo of mass $\mathrm{M_h}$ is
\begin{equation}
\lUV(M_h,z) = \epsilon_*(z) \bigg(\frac{\Omega_b}{\Omega_m}\bigg) M_h(z) L_{1500}.
\end{equation}
Here, $\epsilon_*(z)$ is the star formation efficiency at redshift $z$, denoting the fraction of gas turned into stars in a burst. This is a free parameter that is calibrated by matching to the relatively bright part (i.e. $\muv \lsim -18$) of the observed evolving ultraviolet luminosity function (UV LF) at $z \sim 5-10$. Finally, 
$L_{1500} = 10^{33.07} ~ [{\rm erg\, s^{-1} \, \AA^{-1} \,} \Msun^{-1}]$ is the specific UV luminosity at $\lambda = 1500 \, \mathrm{\AA}$, obtained from the population synthesis code {\small STARBURST99} \citep{leitherer1999} assuming a Salpeter initial mass function \citep[IMF;][]{salpeter1955} between $0.1-100\Msun$, a stellar metallicity value of $5\% \, \zsun$ (where $\zsun$ is the solar metallicity) and an age of 2 Myrs. This age should be treated as an illustrative example. As seen, the UV luminosity is determined by a combination of $L_{1500}$ and $\epsilon_*(z)$. Therefore, a decrease in the value of $L_{1500}$ \citep[see Eqn. 14,][]{dayal2014a} when assuming an age of a hundred million years for the dominant stellar population \citep[see e.g.][]{laporte2021} could be countered by scaling up $\epsilon_*(z)$ by the same amount ($\sim 50$). We also note that this ``first-order" calculation ignores the impact of feedback (both from supernovae and reionization) that is crucial in determining the properties of low-mass galaxies \citep{sobacchi2013b,dayal2014a,hutter2021}, as well as the impact of dust attenuation, which is important in determining the visibility of high-mass galaxies \citep[e.g.][]{dayal2010a,bowler2015,bouwens2021b}. 

\begin{table}
\centering
\caption {For $z=5-10$ we show the star formation efficiency required to match to the observed UV LF.}
\begin{tabular}{|c|c|c|c|c|c|c|}
\hline
  $z$ & 5 & 6 & 7 & 8 & 9 & 10 \\
  \hline
  $\log(\epsilon_*)$ & $-2.8$ & $-2.7$ & $-2.55$ & $-2.45$ & $-2.3$  & $-2.3$  \\
  \hline
 \end{tabular}
 \label{tab:sfe}
\end{table} 

As shown in Fig. \ref{fig:uvlf}, these simple assumptions result in a theoretical UV LF that, within error bars, is in quite good agreement with observations over $\muv$ values ranging between $-18$ to $-23.5$ at $z \sim 5-9$. The corresponding values of $\epsilon_*(z)$ are shown in Table \ref{tab:sfe}. Between $z \sim 5-9$, $\epsilon_*$ roughly evolves with redshift as
\begin{equation}
\log[\epsilon_*(z)] = 0.12z - 3.4 \, .
\label{sfe_fnz}
\end{equation}

Although the star formation efficiency seems to saturate between $z \sim 9$ to $z \sim 10$, this must be treated with caution given the paucity of data at $z \sim 10$. We therefore use two limiting cases to calculate the UV LF at $z \sim 13$: (1) $\log (\epsilon_*)$ saturates at a value of $-2.3$ (i.e., a star formation efficiency of $0.5\%$) at $z \gtrsim 9$; (2) we assume that the redshift trend of the star formation efficiency inferred in Eqn. \ref{sfe_fnz} holds also at $z \gtrsim 10$.

We now compare in Fig. \ref{fig:uvlf} the results of such scaling to the observational UV LF at $z \sim 13$. As seen, \citet{harikane2022} infer a number density of $\sim 10^{-7.4} {\rm cMpc^{-3}}$ associated with their two sources, centered at $\muv \sim -23.5$. However, at the same number density, our UV LFs predict galaxies to be much fainter, with $\muv \sim -20.4$ ($\sim -21.7$) for case 1 (2). Indeed, the theoretical scalings, which reproduce the observed galaxy populations well at $z \sim 5-10$, are unable to produce any bright galaxies with $\muv \sim -23$ at $z \geq 9$, including the sources at $z \sim 13$.

We then use the relation $\mathrm{SFR}_{\rm UV} = \kappa \lUV$ to infer the UV SFR. Assuming continuous star formation for 100 Myrs at $5\% \zsun$ metallicity, we obtain a value of $\kappa = 8.23 \times 10^{-29} \, [\Msun~{\rm yr^{-1} ~erg^{-1} \, s \, Hz}]$ using {\small STARBURST99}. This results in $\mathrm{SFR}_{\rm UV}=78.9 ~\Msun {\rm yr^{-1}}$ and $\mathrm{SFR}_{\rm UV}=125.2 ~ \Msun {\rm yr^{-1}}$ for HD1 and HD2, respectively, as also noted in Table \ref{table1}. On the other hand, our theoretical scalings produce $\mathrm{SFR}_{\rm UV}$ values of about $ 5.3~(16.5) \Msun {\rm yr^{-1}}$ for case 1 (2) for the observed number density. This implies that these galaxies are {\it extreme star formers}, whose UV SFR (and hence star formation efficiencies) are a factor $\sim 15-24$ $(4.8-7.6)$ higher than expected in case 1 (2) for a given halo mass. In order to cross-check our result, we can also look at the halo masses that would host $\mathrm{SFR}_{\rm UV}$ values of the order of $79-125 \Msun {\rm yr^{-1}}$ at $z \sim 13$. Using the scalings above, these correspond to $\mathrm{M_h} \sim 10^{12.17-12.37} ~ (10^{11.6-11.8}) \Msun$ for case 1 (2). However, the number densities of such halos are $\leq 10^{-10.8} {\rm cMpc^{-3}}$, which are three orders of magnitude lower than the observationally inferred values. 

To conclude, the number densities of $z \sim 13$ galaxies inferred by \citet{harikane2022} are too large compared to the evolution expected from lower redshifts ($z \sim 5-9$). Indeed, the observed SFRs are more likely to be hosted in significantly larger halos of $\mathrm{M_h} \geq 10^{11.6}\Msun$, whose number density at $z=13$ would be $\leq 10^{-10.8} {\rm ~cMpc^{-3}}$. Based on the combination of their UV luminosity and number density, some plausible explanations for the observed $z \sim 13$ galaxy candidates are \citep[see also][]{Harikane_2021, harikane2022}: {\it (i)}: these galaxies are {\it extreme star-formers} whose star formation efficiency is $5-24\times$ higher than expected for a given halo mass \citep{Harikane_2021}; {\it (ii)}: at any redshift, the star formation efficiency increases with increasing halo mass and is countered by increasing dust attenuation from $z \sim 10-5$, leading to a lack of evolution in the bright-end of the UV LF \citep[e.g.,][]{bowler2020, Harikane_2021}; {\it(iii)} these galaxies have an IMF that is more top-heavy compared to the Salpeter IMF assumed here \citep{harikane2022}. An extremely top-heavy IMF where the number of stars ($N$) of a given mass ($\rm M_\star$) scales as $N \propto \rm M_\star^{-0.95}$ \citep[][]{fardal2007} could produce about 8 times more UV luminosity per unit star formation rate using, e.g., {\small STARBURST99}.

\section{The Quasar Hypothesis}
\label{sec:quasar}
We then explore the quasar hypothesis, i.e., the possibility that these galaxies host early super-massive black holes, and that the elevated UV is powered by their accretion.
By assuming that the UV luminosity of HD1 and HD2 is  solely generated by accretion power, we can obtain an estimate of the black hole mass and of its accretion rate. 

\begin{figure}
\includegraphics[width=0.975\columnwidth]{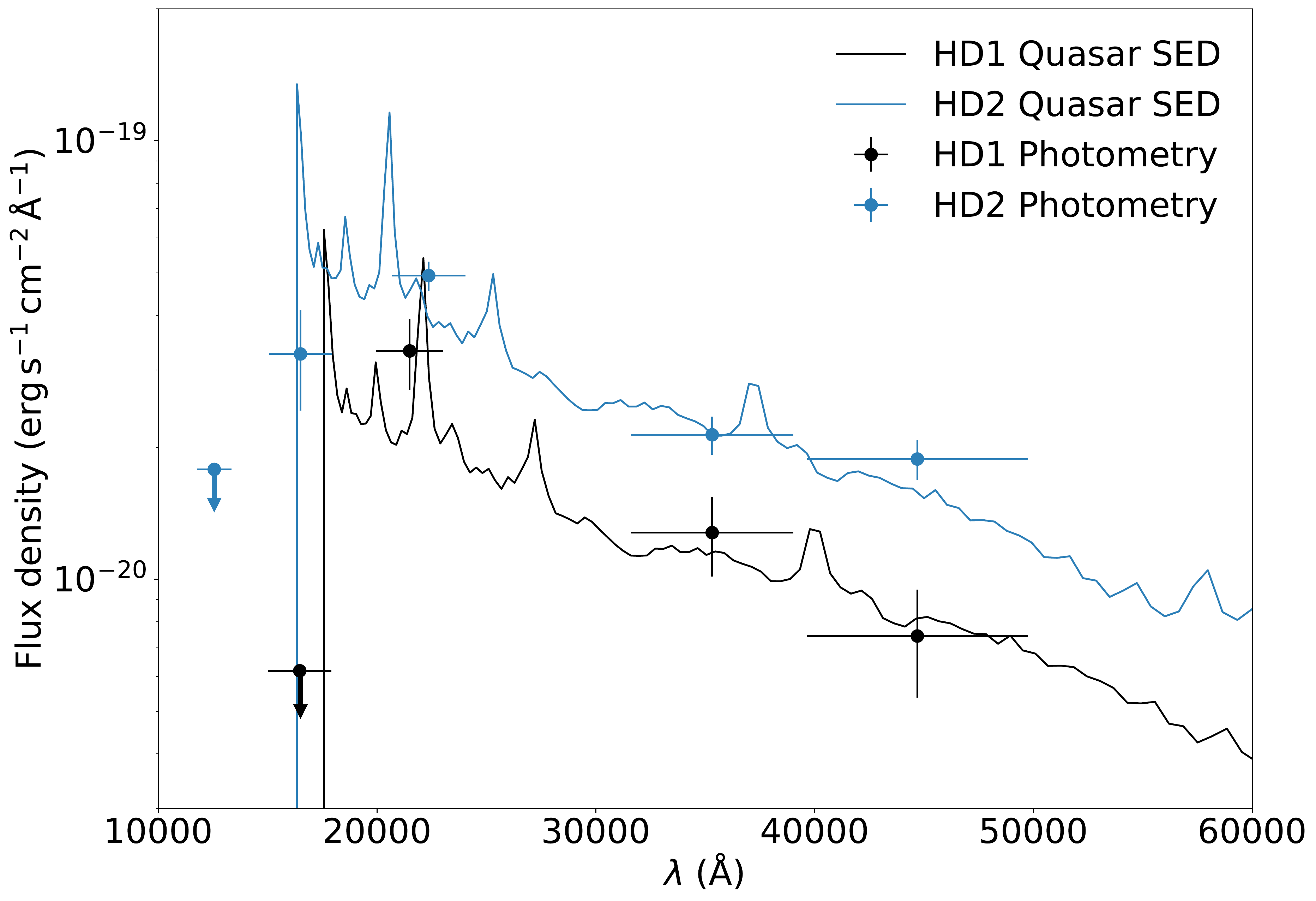}
    \caption{Fitting of photometric points with quasar SEDs. The points and their associated $2\sigma$ uncertainties for sources HD1 and HD2 are from \citet{harikane2022}. Arrows indicate upper limits.}
    \label{fig:BH_SED_fit}
\end{figure}

Following the results of \citet{marconi2004}, we use the standard scaling $\mathrm{L_\nu} \propto \nu^{-0.44}$ to convert the UV luminosity of HD1 and HD2 into a B-band luminosity. We then use the conversion \citep{marconi2004}:
\begin{equation}
    { \frac{\log \lbh}{\nu_B \mathrm{L_{\nu B}}}} = 0.80 -0.067 \gamma  
    + 0.017 \gamma^2 - 0.0023 \gamma^3,
\end{equation}
where $\gamma = {\log}\, \lbh-12$, to infer the bolometric luminosity of the black hole, $\lbh$. The luminosities of the two sources are $\lbh = 1.5\times 10^{46} \, \mathrm{erg \, s^{-1}}$ and $\lbh = 2.1\times 10^{46} \, \mathrm{erg \, s^{-1}}$, respectively.
By assuming that these sources are shining at the Eddington luminosity $L_{\rm Edd} \approx 1.26 \times 10^{38} \, (\mbh/\Msun) \, \mathrm{erg \, s^{-1}}$, we can estimate the associated black hole mass $\mbh$.
For the two sources, the inferred black hole masses are $\mbh = 1.2\times 10^{8} \Msun$ and $\mbh = 1.7\times 10^{8} \, \Msun$, respectively, and their accretion rates are $\dot{M}_{\rm BH} = 2.7 \Msun {\rm yr^{-1}}$ and $\dot{M}_{\rm BH} = 3.7 \Msun {\rm yr^{-1}}$ (assuming a standard matter-to-energy conversion efficiency $\epsilon = 0.1$). These properties are reported in Table \ref{table1}.

\begin{table}
\caption {For HD1 and HD2, we show UV luminosities observed at a rest-frame wavelength of 1500\AA ($\lUV$ in units of $L_\odot$) and the associated UV SFR (${\rm SFR_{UV}}$ in units of $M_\odot {\rm yr^{-1}}$). If all of this luminosity is attributed to black hole accretion, the associated bolometric luminosity ($\lbh $ in units of ${\rm erg ~ s^{-1}}$) and black hole mass ($\mbh$, in $M_\odot$) are also reported.}
\centering
\begin{tabular}{ccccc}
\hline
Source & $\lUV$ & ${\rm SFR_{UV}}$ & $\lbh$ & $\mbh $                            
\\ \hline
HD1 & $10^{11.7}$ & 78.9 & $1.5 \times 10^{46}$ & $1.2 \times 10^8$ \\
HD2 & $10^{11.9}$ & 125.2 & $2.1\times 10^{46}$ & $1.7 \times 10^8$ \\
\hline
\end{tabular}
 \label{table1}
\end{table}

Note that these estimated masses are $1-2$ orders of magnitude lower than those of the most massive black holes at $z \sim 7$, which, however, had $\sim 440$ Myr more time to grow (see, e.g., \citealt{Inayoshi_review_2019, Pacucci_2022}). Also, note that the estimated bolometric luminosities are the same as that of a low-luminosity quasar detected at $z=7.07$ \citep{Matsuoka_2019a}.
With a bolometric luminosity of $\sim 2\times 10^{46} \, \mathrm{erg \, s^{-1}}$, assuming $z = 13$, these sources have a bolometric flux of $\approx 10^{-14} \, \mathrm{erg \, s^{-1} \, cm^{-2}}$. We assume an X-ray bolometric correction of $\sim 10\%$, given that the observed frame $0.5 - 10$ keV corresponds to the rest-frame $7 - 140$ keV, which should not be affected by significant obscuration, due to the very high energies involved.
Hence, we obtain an estimated X-ray flux of $\approx 10^{-15}\, \mathrm{erg \, s^{-1} \, cm^{-2}}$. Unfortunately, both HD1 and HD2 are very close to the border of the field of view in the COSMOS and UDS fields, where the detection of point-like sources is particularly challenging. Targeted, deep X-ray observations could ultimately clarify the nature of these sources, as the X-ray flux from a starburst is lower than that expected from a black hole.

Having obtained an estimate of the black hole mass and bolometric luminosity, we proceed to check if the observed photometry can be modeled by standard spectral energy distributions (SEDs) for quasars.
We caution the reader that our intent here is not to obtain a best-fit SED given the photometric data, because the estimate of the black hole mass and of the bolometric luminosity are very tentative. Our goal is to use standard black hole SEDs to show that the quasar hypothesis is compatible with the photometric data.

We use the standard atlas of quasar SEDs by \cite{Shang_atlas_2011}, a modern version of the classic catalog by \cite{Elvis_1994}. In our range of interest ($1200 \, \mathrm{\AA}$ to $4300 \, \mathrm{\AA}$ rest-frame) the two SED templates provided (radio-loud vs radio-quiet) are similar to each other \citep{Shang_atlas_2011}.
We choose the quasar 3C 263 to normalize the SED template, a well studied source with plenty of available data regarding accretion rates and black hole mass. In addition, \cite{Shang_atlas_2011} discuss how the radio-loud SED is constructed from higher-redshift sources, when compared to the radio-quiet one.
We obtain that the mean flux density of 3C 263 at $\sim 1060 \, \mathrm{\AA}$ (observed frame) is $1.19 \times 10^{-14} \, \mathrm{erg \, s^{-1} \, cm^{-2} \, \AA^{-1}}$ \citep{Shang_atlas_2011}.
Furthermore, we need to rescale it taking into account the different physical properties of 3C 263 when compared with the estimated properties of HD1 and HD2. The black hole mass of 3C 263, at $z=0.646$, is $10^{9.1} \Msun$, with a mass accretion rate of $2.95 \, \Msun \, \mathrm{yr^{-1}}$, leading to an Eddington ratio $f_{\rm Edd} = \dot{M}_{\rm BH}/\dot{M}_{\rm Edd} \sim 0.11$ \citep{Mc_Lure_2006, Daly_2021}.
The normalization of the SED template is based on the definition of flux density such that $F_{\lambda} = L_{\lambda}(4\pi D_L^2)^{-1} (1+z)^{-1}$ where $L_{\lambda}$ is the luminosity density and $D_L$ is the luminosity distance.
Furthermore, in the specific case of a black hole, $L \propto \fedd \, \dot{M}_{\rm Edd} \propto \fedd \, \mbh$.

Regarding the properties of the putative $z\sim 13$ sources, they are assumed as follows: HD1 is black hole with mass $\sim 1.2 \times 10^8 \Msun$ at $z = 13.3$, while HD2 is black hole with mass $\sim 1.7 \times 10^8 \Msun$ at $z = 12.3$.
Finally the SED template is truncated for wavelengths shorter than the Ly$\alpha$, at $1215.67 \, (1 + z) \, \mathrm{\AA}$ with an exponential decay factor $\exp{(-\tau)}$, where $\tau$ is calculated following \cite{Madau_1995}. 

The SEDs are a good fit to the photometric points with Eddington ratios $f_{\rm Edd} \approx 1.0$ for HD1 and $f_{\rm Edd} \approx 1.1$ for HD2 (see Fig. \ref{fig:BH_SED_fit}). Note that the actual Eddington ratios estimated from photometric data are similar to the initial working assumption $f_{\rm Edd} = 1$ used in the quasar hypothesis.

The flux densities estimated for HD1 and HD2 are roughly in accordance with the ones reported in \cite{Wang_2021} for the farthest confirmed quasar detected thus far, considering that J0313--1806 is almost exactly $10$ times more massive than HD2, has an Eddington ratio $\sim 1.6$ times lower than HD2, and it is at redshift $z = 7.642$ instead of $z = 12.3$.

\section{Constraints on Black Hole Growth if the Sources are Quasars}
\label{sec:growth}
The growth of a $\sim 10^8 \Msun$ black hole by $z \sim 13$ is very challenging, but not impossible given current growth models (see, e.g., \citealt{Woods_2019,  Inayoshi_review_2019}). The age of the Universe at $z = 13$ is $330$ Myr, which is about $\sim 350$ Myr earlier than J0313--1806, the current record-holder as the farthest quasar \citep{Wang_2021}.

If a heavy black hole seed formed HD1 and HD2 and reached $\sim 10^8 \Msun$ already at $z \sim 13$, then, assuming that the growth rate stays constant, we should be observing super-massive black holes with mass in excess of $\sim 10^{12} \Msun$ by $z \sim 7$. We do not observe such ultra-massive black holes, as the heaviest one detected at high-$z$ is on the $\sim 10^{10} \Msun$ scale \citep{Wu_2015}. This can be explained by: (i) a black hole cannot sustain accretion at the Eddington rate for $\sim 500$ Myr, probably due to the lack of available gas; (ii) even if it grew to masses $\gg 10^{10} \Msun$, it would not be electromagnetically detectable (see, e.g., \citealt{Inayoshi_2016_MaxMass, King_2016}).

Recently, \cite{Pacucci_2022} pointed out that the discovery of farther quasars would improve the estimation of the parameters for black hole growth: Eddington ratio ($f_{\rm Edd}$), duty cycle (${\cal D}$), seed mass ($M_{\rm seed}$) and radiative efficiency ($\epsilon$). This is due to the fact that a higher detection redshift implies a shorter time between seeding and observation.
\cite{Pacucci_2022} calculated that the uncertainties in the determination of the growth parameters decrease by $\sim 5$ when the detection redshift goes from $z=9$ to $z=12$.

While the interested reader is referred to \cite{Pacucci_2022} for an in-depth description of the model, we provide some essential details here. This technique of ``blind estimation'' is aimed at constraining the entire parameter space $[f_{\rm Edd}, {\cal D}, \epsilon, M_{\rm seed}]$ that could allow the growth of a given black hole by the observed redshift. For this experiment, we assume a seeding redshift $z = 25$. Additionally, we assume the following flat priors on the growth parameters: (i) $\fedd \in [0,1]$; (ii) ${\cal D} \in [0,1]$; (iii) ${\epsilon} \in [0.057,0.32]$ (within the thin disk efficiency regime, see, e.g., \citealt{Bardeen_1970, Fabian_2019}); (iv) light seeds: $\mathrm{Log_{10}} {M_{\rm \bullet, L}} \in [0,3]$, heavy seeds: $\mathrm{Log_{10}}{M_{\rm \bullet, H}} \in [3,6]$.

\begin{figure}
\includegraphics[width=\columnwidth]{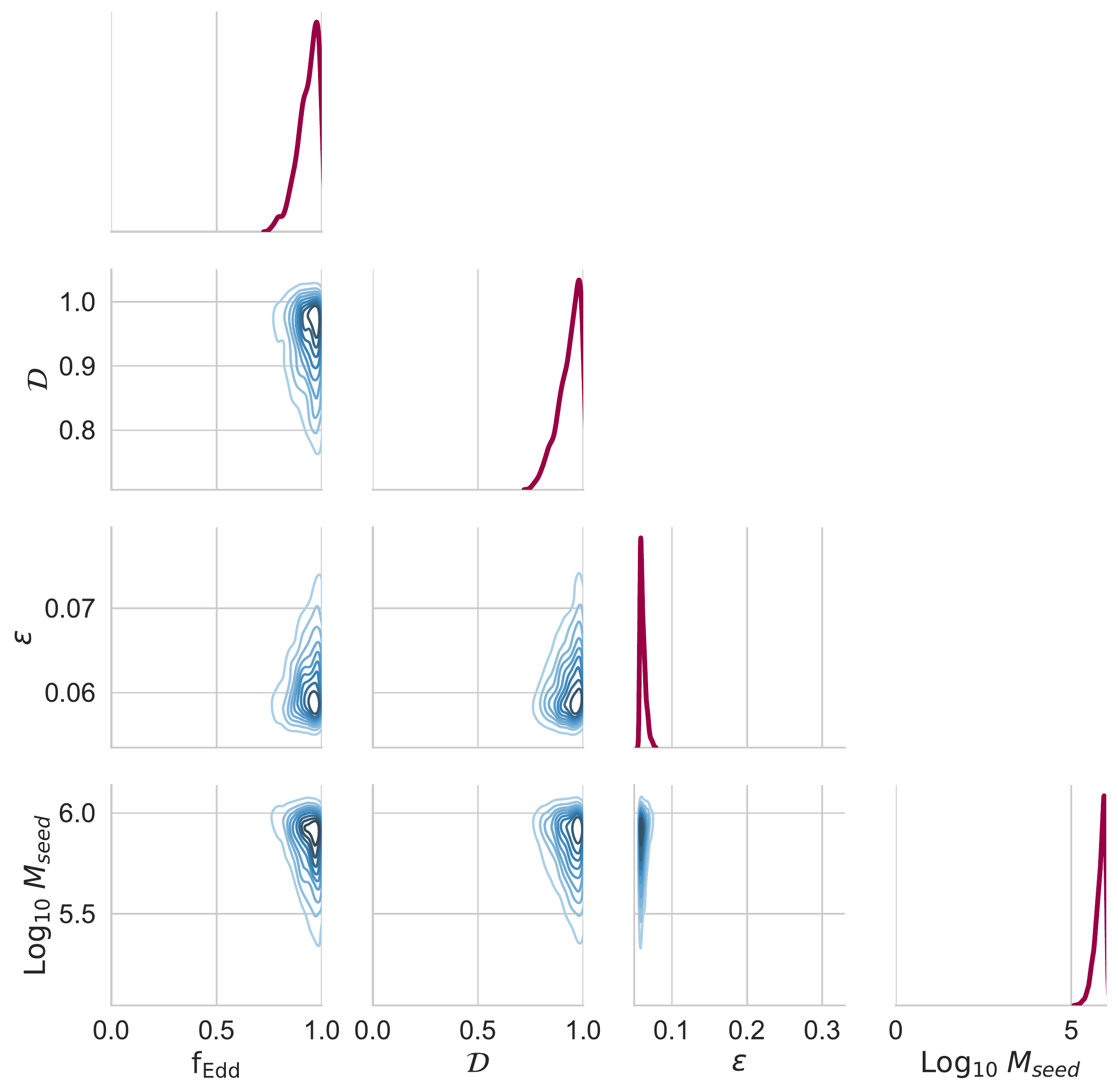}
    \caption{Growth parameters constrained from the observation of HD1. The method applied to HD2 yields a very similar corner plot and constraints on parameters \citep{Pacucci_2022}.}
    \label{fig:BH_growth}
\end{figure}

Applying the same technique, we present our results for the putative black hole in HD1 in Fig. \ref{fig:BH_growth}. The 2D joint distributions in the lower half of the panel show contour plots, where levels are iso-proportions of the probability mass function with each partition comprising $10\%$ of it. Our analysis shows that a detection of a $1.2 \times 10^8 \Msun$ black hole by $z = 13.3$ (the estimated parameters for HD1) would require mean Eddington ratios and duty cycles above $\sim 0.7$, very low accretion efficiencies ($\epsilon \sim 6\%$, typical of no-spin conditions) and a black hole seed mass well into the heavy seeds regime ($M_{\rm seed} \gtrsim 10^5 \Msun$).

The preferred values for the growth parameters for HD1, presented as mean $\pm$ standard deviation, are: $f_{\rm Edd} = 0.93 \pm 0.05$, ${\cal D} = 0.93 \pm 0.05$, $\epsilon = 0.06 \pm 0.003$ and $\log_{10}M_{\rm seed} = 5.82 \pm 0.14$. Note that these constraints are similar to what obtained for a $10^9 \Msun$ quasar detected at $z=12$ \citep{Pacucci_2022}. Additionally, the number density at $z\sim 13$ estimated from these two sources in \cite{harikane2022} is $4\times 10^{-8} \, \mathrm{mag^{-1} \, Mpc^{-3}}$, with uncertainties of a factor $\sim 50$. Interestingly, \cite{Matsuoka_2018} estimates a quasar number density of $2 \times 10^{-8} \, \mathrm{mag^{-1} \, Mpc^{-3}}$ at $z=6$ for sources of the same UV absolute magnitude. Note that the number density of halos with mass $\mathrm{M_h} < 10^{10} \Msun$ is similar at $z=10$ and $z=6$, as their ratio is $<10$ \citep{Loeb_2010}. If the black hole mass is dictated by halo velocity dispersion, then the number density of black holes $\sim 10^8 \Msun$ would be similar at $z \sim 6$ and $z \sim 13$, further supporting the quasar hypothesis.

\section{Conclusions and discussion}
\label{sec:conc}
Motivated by the detection of two $z \sim 13$ galaxy candidates \citep{harikane2022}, in this Letter we discuss their physical nature.
In particular, we present two fiducial models to explain the observed UV emission from HD1 and HD2 in the case that: (i) they are ultra UV-bright star-forming galaxies at $z\sim 13$; or (ii) they are quasars at $z\sim 13$. Alternatively they could be $z\sim 4$ passive galaxies, despite the fit obtained against the measured photometry being less significant in this case \citep{harikane2022}. Our key findings are:
\begin{itemize}
    \item The combination of the observed UV luminosity ($\muv \sim -23.5$) and number density ($\sim 10^{-7.4} {\rm ~cMpc^{-3}}$) cannot be reproduced when using averaged trends for star-forming galaxies extrapolated from lower-redshifts. A few solutions are that these galaxies are possibly either extreme star-formers (with star formation efficiencies $5-24\times$ higher than expected from average relations), or have a star formation efficiency that increases with halo mass, or have a top-heavy IMF.
    \item The quasar hypothesis is feasible, as we have shown that the observed UV luminosity could be produced by a $\sim 10^8 \Msun$ black hole, accreting at or slightly above the Eddington rate ($f_{\rm Edd} \sim 1.0-1.1$) at $z\sim 13$.
    \item The black hole model would require heavy seeds and large mean Eddington ratios and duty cycles (assuming a standard thin-disk accretion model) to be feasible. Additionally, its growth would need to be stunted by lack of gas in the period $z=13-7$ to avoid $10^{12}\Msun$ quasars at $z\sim 7$. Alternatively, such ultra-massive black holes might become electromagnetically undetectable.
\end{itemize}

More probable than these ``extreme scenarios" is a situation in which the UV luminosity of these $z \sim 13$ sources is contributed by a combination of star formation and black hole accretion. Indeed, at $z \sim 5-6$, they contribute almost equally to the UV luminosity for $\muv \sim -23.5$ galaxies \citep{ono2018, piana2022,Harikane_2021}. The contribution from star formation would decrease the predicted mass of the black holes, requiring less extreme growth conditions to assemble by $z \sim 13$.

If the $z\sim 13$ nature of these sources is confirmed by further spectroscopic observations, and the black hole hypothesis holds, these sources will provide a remarkable laboratory to test seed formation models. As explained in \S \ref{sec:growth}, a black hole of $\sim 10^8 \Msun$ by $z \sim 13$ would likely require a heavy seed, with typical mass $\gtrsim 10^5 \Msun$. Many formation channels have been proposed to form such seeds already at very high-$z$ (see, e.g., \citealt{Woods_2019, Inayoshi_review_2019}). Alternatively, heavy seeds could be formed very early in the history of the Universe as primordial black holes (PBHs) with a mass scale $M_{\rm PBH} \sim 10^5 \, (t/\mathrm{1 \, s}) \, \Msun$, where $t$ is the formation time after the Big Bang (see, e.g., \citealt{Carr_2018, Inayoshi_review_2019}). The formation of very massive seeds is suppressed, and the most-likely mass formed is close to the Chandrasekhar mass. Despite this, the formation of $\sim 10^{4-5} \Msun$ seeds remains possible, with recent papers exploring the possibility that PBHs could play an important role in the formation of the first super-massive black holes (see, e.g., \citealt{Hasinger_2020, Cappelluti_2022}). Additional data, from forthcoming facilities such as the JWST, the Extremely Large Telescope (ELT) and the NGRST, will be crucial in answering many of these open questions and shedding light on galaxy formation at these previously inaccessible epochs.

\vspace{-0.7cm}
\section*{Acknowledgments} 
F.P. acknowledges support from a Clay Fellowship administered by the Smithsonian Astrophysical Observatory.
P.D. acknowledges support from the ERC starting grant StG-717001 (``DELPHI"), from the NWO grant 016.VIDI.189.162 (``ODIN") and the EC and University of Groningen's CO-FUND Rosalind Franklin program. Y.H. is supported by the JSPS KAKENHI Grant (21K13953). A.K.I. is supported by NAOJ ALMA Scientific Research Grant Numbers 2020-16B. This work was also supported by the Black Hole Initiative at Harvard University, funded by grants from the John Templeton Foundation and the Gordon and Betty Moore Foundation.
F.P. thanks fruitful discussions with Ramesh Narayan and P.D. thanks Anupam Mazumdar for insightful comments.

\vspace{-0.6cm}
\section*{Data Availability}
Data generated in this research will be shared on reasonable request to the corresponding author.

\vspace{-0.6cm}
\bibliographystyle{mnras}
\bibliography{mybib,ms}

\label{lastpage} 
\end{document}